\documentclass[prc,aps,floatfix,nofootinbib,showpacs,amsmath,amssymb]{revtex4}

\usepackage{graphicx}
\usepackage{epsfig}
\usepackage{dcolumn}
\usepackage{bm}

\def\beq{\begin{equation}}
\def\eeq{\end{equation}}

\newcommand{\rv}{\mathbf{r}}

\newcommand{\kv}{\mathbf{k}}
\newcommand{\qv}{\mathbf{q}}

\newcommand{\be}{\begin{eqnarray}}
\newcommand{\ee}{\end{eqnarray}}
\newcommand{\ba}{\begin{array}}
\newcommand{\ea}{\end{array}}

\newcommand{\een}{\nonumber\end{eqnarray}}
\def\v#1{{\bf#1}}
\def\bra{\langle}
\def\ket{\rangle}
\def\kgs{\vert \phi_0 \rangle}
\def\bgs{\langle \phi_0 \vert}
\def\egu{\, =\, }

\def\cap{\noindent}

\def\le#1{\label{eq:#1}}
\def\re#1{\ref{eq:#1}}

\def\creas{{}^\dagger}

\def\v#1{{\bf#1}}

\begin{document}
\thispagestyle{empty}
\vspace*{0.5 cm}
\title{Correlations in Nuclear Matter}
\author{M. Baldo $^1$\footnote{ email: baldo@ct.infn.it } and H.R. Moshfegh $^{1,2}$ \footnote{ email: hmoshfegh@ut.ac.ir }}
\affiliation{$^1$INFN, Sezione di Catania
 Via S. Sofia 64, I-95123, Catania, Italy \\
$^2$Department of Physics, University of Tehran,\\ P.O.B. 14395-547, Tehran, Iran } 
\begin{abstract} We analyze the nuclear matter correlation properties in terms of the pair correlation function. To this aim we systematically compare the results for the variational method in the Lowest Order Constrained Variational (LOCV) approximation and for the Bruekner-Hartree-Fock (BHF) scheme. A formal link between the Jastrow correlation factor of LOCV and the Defect Function (DF) of BHF is established and it is shown under which conditions and approximations the two approaches are equivalent.
From the numerical comparison it turns out that the two correlation functions are quite close, which
indicates in particular that the DF is approximately local and momentum independent. The Equations of State (EOS) of Nuclear Matter in the two approaches are also compared. It is found that once the three-body forces (TBF) are introduced the two EOS are fairly close, while the agreement between the correlation functions holds with or without TBF. 
\end{abstract}
 \vskip 0.3 cm
%

\maketitle

\section{Introduction}
The structure and properties of nuclear matter is one of the central issues in the development of nuclear many-body
theory. Nuclear matter is of great relevance for the physics of supernova, neutron stars and heavy ion collisions,
for the development of density functionals in nuclear structure studies and for the understanding at fundamental
level of the low energy baryon-baryon interaction. For a review see reference \cite{report}. Different many-body
theories \cite{book,Navarro} have been developed to approach this problem. One can mention the variational method \cite{V1,V2,V3,V4,V5,V6,V7,V8}, the Monte-Carlo method in its different versions
\cite{MC1,MC2,MC3,MC4,MC5,MC6} and the diagrammatic expansion methods, in particular the Brueckner-Bethe-Goldstone hole-line expansion \cite{book} and the Self Consistent Green Function scheme \cite{S1,S2,S3,S4,S5,S6,S7,S8}. One of the main goal of this effort along the
years has been the explanation of the saturation point of nuclear matter that can be extracted phenomenologically through various experimental method, in particular the analysis of the binding energy of nuclei and of the electron elastic scattering cross sections \cite{report}. However, besides the saturation point, one of the most
important characteristics of nuclear matter is its correlation structure. In fact the presence of a hard core in
the nucleon-nucleon interaction produces a correlation "hole" between two nucleons that can be described by the
correlation function. The latter is also determined by the intermediate and long-range interaction, typical of the nuclear two-nucleon potential. The correlation function is a key quantity to characterize each many-body scheme
and to understand the corresponding numerical results. 
In scattering studies, the spectral function of many fermionic system gives the important quantities of interests and the short and long range correlation functions are very important factors for calculating the spectral functions  \cite{Benh}. The connection of the correlation function and the spectral function
is not straightforward, but it has been elucidated in ref. \cite{claudio} in the Brueckner-Hartree-Fock (BHF) framework  
and in ref. \cite{Modar12} for the Lowest Order Constrained Variational (LOCV) method, where it is particularly transparent.  
Furthermore several phenomena that occur in
neutron star matter are closely linked to the correlation function, like e.g. dissipation due to shear viscosity
and neutrino transport.
It appears then natural to look for a comparison between the correlation functions from
different many-body schemes. In this paper we present a detailed comparison between the Bethe-Brueckner-Goldstone
(BBG) method \cite{book} and the variational method, as developed within the LOCV framework. Both methods have been applied systematically to nuclear matter with different two-body
interactions. The results for the saturation point and other physical parameters, like the compressibility at high
density \cite{rep_jpg,mod_jpg}, the critical temperature of the liquid-gas phase transition \cite{lidia,lg_locv}, are close but not completely in agreement.
One of the main goals of this work is to present an analysis of the correlation function that could help understanding the reason of the agreements and the discrepancies by the comparison of the corresponding correlation properties.

\section{The variational method \label{var}}
The method of Lowest Order Constrained Variational approach is among the microscopic methods that were developed to calculate
the bulk properties of homogeneous nuclear fluids such as the saturation quantities by using the realistic nucleon-nucleon interaction i.e. Reid68 and $\Delta-$Reid (the modified Reid potential with inclusion of isobar degrees of freedom) \cite{Mod79}. This method was reformulated to include more sophisticated interactions \cite{Bord} such as
$UV_{14}, AV_{18}$ \cite{wir} and charge dependent Reid potential (Reid93)\, \cite{Reid}. The LOCV method has
been also developed for calculating the various thermodynamic properties of hot and frozen homogeneous fermionic
fluids such as symmetric and asymmetric nuclear matter \cite{Hamid1}, $\beta-$stable matter \cite{Hamid2}, Helium
3 \cite{Hamid3}, electron fluid \cite{elec} with different realistic interactions. Recently the LOCV formalism was
developed for covering the relativistic Hamiltonian with a potential which has been fitted relativistically to
nucleon-nucleon phase shifts \cite{rel}. The LOCV calculation is a fully self-consistent technique with state
dependent correlation functions. There is no free parameter in this method, except those included in the
interactions. Considering constraint as in the form of normalization condition is another advantage of LOCV
formalism. This assumption keeps the higher order terms as small as possible and it also assumes a particular form
for the long range behavior of the correlation functions in order to perform an exact functional minimization of
the two-body energy with respect to the short range parts of correlation functions. The functional minimization
procedure represents an enormous computational simplification over the unconstrained methods, where the short
range behavior of the correlation functions is parametrized, that attempt to go beyond the lowest order
\cite{beyond}. To test the convergence of the LOCV method for nuclear matter and helium 3, the calculations were
performed beyond the lowest order and the three-body cluster energy was evaluated with both the state averaged and
state dependent correlation functions \cite{TBF}. The smallness of the normalization (the convergence parameter) and of
the three body cluster energy indicated that at least up to the twice empirical nuclear matter saturation density,
the cluster expansion converges reasonably and stopping after two-body cluster terms is a fair approximation.
   In the LOCV method, we use an ideal Fermi gas type wave functions, $\phi_{i}$ , for the single particle states
   and we employ the variational techniques to find the wave function of interacting system \cite{Mod79}-\cite{Reid},i.e,
\beq
\Psi=F \, \Phi,
\eeq 
where $\Phi$ is the uncorrelated Fermi system wave function (Slater determinant of plane waves) and the factor $F(1,2..,A)$ is the many-body correlation function, defined as product of two body correlation functions $f(i,j)$ (Jastrow form) and assumes that they are operators,
\beq
F={\cal S} \prod_{i<j}f(i,j),
\eeq
where $\cal S$ is a symmetrizing operator. 
   The many-body energy term $E[f]$, which is a functional of the $f$' s, is calculated by constructing a cluster expansion
   for the expectation value of Hamiltonian $H$ of the system.
\beq
E[f] \,=\, \frac{1}{A}\, \frac{<\, \Psi\, |\, H\, |\, \Psi\, >}{<\, \Psi\, |\, \Psi\, >} \,=\, E_1 \,+\, E_2 \,
+\, \cdots\cdots \,>\, E_0.
\eeq
 where $E_0$ is the true ground state energy and A is the particle number.
 In the lowest order we truncate the above series after $E_2$ i.e. two-body energy.
 The one body term $E_1$ is independent of the f and is just the familiar Fermi gas kinetic energy.
The two-body energy term is defined as,
\be
E_2 &\,=\,\,& \frac{1}{2A} \sum_{ij} <\, ij\, |\, W\, |\, ij\,>_a  \ \ \ \ ; \ \ \ \ | \, ij\, >_a \,=\, | \, ij\,
> \,-\, | \, ji\, > \nonumber\\
 &\, &  \nonumber\\
W &\,=\, & -\, \frac{\hbar^2}{2m} \big[ f(1,2) , [ \nabla^2 , f(1,2) ] \big] \,+\, f(1,2) V(1,2) f(1,2),
\label{eq:W}\ee
and the two-body anti-symmetrized matrix element $ <\, ij\, |\, W\, |\, ij\,>_a $ are taken with respect to the
single-particle functions composing $\phi_{i}$ i.e. plane-waves. By inserting a complete set of two-particle state
twice in above equation and performing some algebra we can rewrite the two-body term as a functional of
correlation functions \cite{Mod79,Bord,Reid} . In this equation $V(1,2)$ is phenomenological nucleon-nucleon potential such as Reid
type, $UV_{14}$ and $AV_{18}$. At this stage, we can minimize the two-body energy  with respect to the variations
of the correlation functions \cite{Mod79,Bord,Reid} , but subject to the normalization constraint \cite{Mod79}-\cite{rel} :
\beq
 \frac{1}{A} <\, ij\, |\, h^2(1,2) \,-\, f^2(1,2)\, |\, ij\, >_a \,\,=\,\, 1.
\label{eq:norm}\eeq
\noindent The function $h(12)$ is the modified Pauli function, which for the symmetrical nuclear matter take the
following form:
\beq
 h(1,2) \,=\, \Big( \, 1 \,-\, \frac{9}{4} \big(\, \frac{j_1(r_{12})}{r_{12}}\, \big)^2 \Big)^{-\frac{1}{2}},
\eeq
\noindent where $j_1(r_{12})$ is the well known spherical Bessel function of order 1. Note that $[\chi=<\Psi|\Psi>-1]$ plays
 the role of a smallness parameter in the cluster expansion. The above constraint
introduces a Lagrange multiplier through which all the correlation functions are coupled. Then we can write sets
of uncoupled and coupled Euler-Lagrange differential equations with respect to the correlation functions. The
constraint is incorporated by solving these Euler-Lagrange equations only up to a certain distance where the
logarithmic derivative of correlation functions matches those of Pauli function and then we set the correlation
functions equal to Pauli function . As we pointed out before, there is no free parameter in our LOCV formalism
i.e. the healing distance is determined directly by the constraint and the initial conditions.
\section{The BBG expansion \label{BBG}}
\begin{figure} [h]
\begin{center}
\vskip 0.6 cm
\includegraphics[bb= 140 0 300 790,angle=90,scale=0.4]{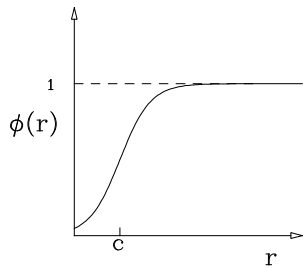}
\vspace{0.3 cm}
   \caption{Schematic representation of the expected effect
  of the core repulsion on the two-body wave function in nuclear matter.}
    \label{fig:fig_corr}
\end{center}
\end{figure}

One of the most known and used microscopic many-body approach to the theory of nuclear matter is the
Bethe-Brueckner-Goldstone (BBG) expansion \cite{book}. In this scheme the original nucleon-nucleon interaction is
systematically replaced by the so-called G-matrix, that describes the two-nucleon scattering amplitude inside the
medium. A modified perturbative expansion is then developed in terms of this effective interaction and the
different terms can be represented by diagrams. The G-matrix can be defined also for singular interaction, e.g.
with a hard core, and it is expected to be "smaller" than the original NN interaction. Although all modern
realistic NN interactions introduce a finite repulsive core, it is however quite large, and therefore in any case
a straightforward perturbative expansion cannot be applied. As discussed in the presentation of the variational
method, the repulsive core is expected to modify strongly the ground state wave function whenever the coordinates
of two particles approach each other at a separation distance smaller than the core radius $c$. In such a
situation the wave function should be sharply decreasing with the two particle distance. The ``wave function'' of
two particles in the unperturbed ground state $\phi_0$ can be defined as ($k_1 , k_2 \leq k_F$) \beq
 \phi(r_1,r_2) \egu \bgs \psi{ }^\dagger_{\xi_1} (\v{r_1})
 \psi{ }^\dagger_{\xi_2} (\v{r_2}) a_{k_1} a_{k_2} \kgs \egu
 e^{i (\v{k_1} + \v{k_2})\cdot \v{R} }
 e^{i (\v{k_1} - \v{k_2})\cdot \v{r} /2 } \ \ ,
\le{deff} \eeq \cap where $\xi_1 \neq \xi_2$ are spin-isospin variables, and $\v{R} = (\v{r_1} + \v{r_2})/2$,
$\v{r} = (\v{r_1} - \v{r_2})$ are the center of mass and relative coordinate of the two particles respectively.
Therefore the wave function of the relative motion in the $s$-wave is proportional to the spherical Bessel
function of order zero $j_0 (k r)$, with $k$ the modulus of the relative momentum vector $\v{k} = (\v{k_1} -
\v{k_2})/2$. The core repulsion is expected to act mainly in the $s$-wave, since it is short range, and therefore
this behavior must be strongly modified. In the simple case of $k = 0$ the free wave function $j_0 (k r)
\rightarrow 1$, and schematically one can expect a modification, due to the core, as depicted in Fig.
\ref{fig:fig_corr}.
\cap The main effect of the core is to ``deplete'' the wave function close to $r = 0$, in a region of the order of
the core radius $c$. Of course, the attractive part of the interaction will modify this simple picture at $r > c$.
If the core interaction is the strongest one, then the average probability $p$ for two particles to be at distance
$r < c$ would be a measure of the overall strength of the interaction. If $p$ is small, then one can try to expand
the total energy shift $\Delta E$ due to the interaction in power of $p$. The power $p^n$ has, in fact, the
meaning of probability for $n$ particles to be all at a relative distance less than $c$.
In a very rough estimate  $p$ is given by the ratio between the volume occupied by the core and the average
available volume per particle \beq
 p \, \approx \, \left( {c\over d} \right)^3,
\le{wound} \eeq \cap with ${4\pi\over 3} d^3 = \rho^{-1}$. From Eq. (\re{wound}) one gets $p \approx {8\over 9\pi}
(k_F c)^3$, which is small at saturation, $k_F = 1.36 \, fm^{-1}$, and the commonly adopted value for the core is
$c = 0.4 \, fm$. The parameter remains small up to few times the saturation density.\par The terms of the
expansion can now be ordered according to the order of the correlations they describe, i.e. the power in $p$ they
are associated with. It is easy to recognize that this is physically equivalent to grouping the diagrams according
to the number of hole lines they contain, where $n$ hole lines correspond to $n$-body correlations. In fact, an
irreducible diagram with $n$ hole lines describes a process in which $n$ particles are excited from the Fermi sea
and scatter in some way above the Fermi sea. Equivalently, all the diagrams with $n$ hole lines describe the
effect of clusters of $n$ particles, and therefore the arrangement of the expansion for increasing number of hole
lines is called alternatively ``hole expansion'' or ``cluster expansion''. For a pedagogical introduction to the
BBG expansion see refs. \cite{book,jpg}, where references to more technical reviews can be found. In ref.
\cite{jpg} the connection of BBG and the variational method is discussed. The relation between the two approaches turns out to be more transparent if the BBG expansion is reformulated in terms of the coupled cluster method (or $e^S$ method) \cite{Kum}. According to this scheme the wave function of the
ground state is written
\beq
    \vert \Psi \ket \, =\, e^{\hat{S}} \vert \Phi \ket,
\label{eq:exp}\eeq \noindent
\noindent where $\hat{S} $ is a correlation operator containing a set of n-body terms which produce excitations of
$n$ particles from below to above the Fermi sea. 
 This method has also a variational character, in the sense that the variation is performed not
 on the ground state wave function but on these correlation terms \cite{Kum,Heisen}. 
Then a set of coupled equations is obtained for the $n$-body correlation functions. The expansion of this set of equations in terms of the order of the correlations is equivalent to a re-ordering of the hole-line expansion in the BBG theory \cite{DayV}. At the $2$-body level of approximation the method is equivalent to the so-called Brueckner      
approximation \cite{DayV,jpg} in the BBG hole-line expansion, and the operator $\hat{S}$ reduces to a two-body operator $\hat{S}_2$
\beq  \hat{S}_2 \,\,\,\,\,  =\, \sum_{k_1 k_2 , k_1' k_2'}
    \bra k_1' k_2' \vert S_2 \vert k_1 k_2 \ket
    {a}\creas(k_1'){a} \creas(k_2')  \,
    a(k_2) a(k_1),
\label{eq:S} \eeq \noindent where the $k$ ' s label hole state , i.e. inside the Fermi sphere, and the $k'$ ' s
particle states, i.e. outside the Fermi sphere. Each quantity $k$ indicates momentum $\kv$ and spin-isospin quantum
numbers. The function $\hat{S}_2$ is the so called ''defect function" of the Brueckner scheme. It can be written in term of the $G$-matrix
and it is just the difference between the in-medium interacting and non interacting two-body wave functions
\cite{book,jpg}. The different terms of the summation commute with each other and expanding the exponential in Eq.
(\ref{eq:exp}) one gets the product of the correlation operators over all sets of momenta,
\beq
\vert \Psi \ket \, =\, \Pi_{\{k\} } \big[\, 1 \,+\, \sum_{k_1' k_2'} \bra k_1' k_2' \vert S_2 \vert k_1 k_2 \ket
    {a}\creas(k_1'){a} \creas(k_2')  \,
    a(k_2) a(k_1)\, \big]  \vert \Phi \ket ,
\label{eq:prod}\eeq
\noindent where the product is over all disjoint pairs of momenta $k_1 , k_2$, in a given partition of the set of all momenta, in agreement with the Brueckner scheme, which is an independent pair approximation. Higher orders in the expansion vanish because they include powers of annihilation or creation
operators. In the square bracket one can recognize the two-body wave function. After Fourier
transformation to coordinate representation and assuming the defect function to be local and independent of total momentum, this expression acquires
the same form as in the variational method, where the two-body wave function plays the role of the correlation
factors $f(i,j)$. However there are relevant differences with the variational method. First of all the BBG
expansion is not explicitly variational, although, as already mentioned, one can recast the expansion in terms of the $e^S$ scheme, which can be formulated by means of a particular variational procedure \cite{Kum,jpg}. Second, the G-matrix, and
therefore the defect function, is in general highly non-local, which means that the two-body wave function is
dependent also on the initial momenta in the Fermi sea, as well as on the total momentum.  This would imply a correlation factor in integral form
for the variational scheme. Furthermore in BBG expansion one introduces a single particle auxiliary potential, in
order to increase the degree of convergence of the expansion. This potential is usually called Brueckner potential and it is determined with a self-consistent procedure \cite{book}. In the variational method no single particle potential is introduced in the minimization procedure. Of course it is hidden in the mean value of the hamiltonian, but it can be calculated only after the optimal many-body wave
function and energy have been obtained, by adding a tiny fraction of particle to the system \cite{Bob}. Finally in the variational method the correlation function is introduced in the mean value of both the kinetic energy and the interaction term. It is a peculiarity of the BBG expansion that the total energy is written as the sum of the unperturbed kinetic energy and the correlated interaction energy. The latter includes of course implicitly the effect of the correlation on the kinetic energy due to the momentum dependence of the single particle potential and of the G-matrix.
\par
It is one of the main purposes of this work to explore the consequences of these differences on the
correlation properties of the ground state. In turn, the study provides a detailed view of the nuclear matter correlations. Since for both LOCV and the BBG expansion three-body correlations turns out to
be only a fraction of MeV around saturation density \cite{jpg,3hole,3holen}, we restrict the comparison to two-body correlations.  In any case the two-body correlation functions are determined at the BHF level for the BBG expansion and at the two-body Jastrow-like factors for the variational method.  
\section{Formal and numerical comparison}
In order to formulate a meaningful comparison between the two-body microscopic methods, we introduce a mixed representation of the correlation functions. In the expansion of Eq. (\ref{eq:prod}) we separate relative and total momenta and perform the Fourier transformation on the momenta $k'$, i.e. the final ones above the Fermi sea. One gets in this way the correlation function $F$ in coordinate representation, which is dependent on the initial relative momentum and on the total momentum  
\beq
\int\, \frac{d^3 q'}{(2\pi)^3}\, e^{i\qv'\rv}\, \big[\, \delta_{qq'} \,+\, \bra\, q'\, | S_2(P) |\, q\, \ket \big] \,=\, F_B(r; q, P),  
\label{eq:FT}\eeq
\noindent
where the defect function can be written in term of the G-matrix
\beq
\bra\, q'\, | S_2(P) |\, q\, \ket  \,=\, \frac{Q(q',P)}{e(q',q,P)}\bra\, q'\, | G(P) |\, q\, \ket,  
\label{eq:defect}\eeq
\noindent where $Q$ is the average Pauli operator and $e$ is the average two particles excitation energy, see the Appendix for more details. Because of this averaging, the denominator in Eq. (\ref{eq:defect})
can vanish. The integral of Eq. (\ref{eq:FT}) is meant as principal value, in agreement with the BHF calculations of the nuclear matter EOS.\par
The correlation function can be expanded in partial waves and one can define a correlation function for each two-body channel, identified by the quantum numbers $lSJT$ of the relative angular momentum, total spin, total angular momentum and total isospin, respectively. As shown in the appendix, the correlation function $F_B$ has to be compared with the corresponding correlation function $F_V$ for the variational method 
\beq
F_V(r,q) \,=\, f(r)\,\cdot j_l (qr),
\label{eq:fv}\eeq
\noindent where $j_l$ is the spherical Bessel function of order $l$ and $f(r)$ is the correlation function of e.g. Eq. (\ref{eq:norm}). It is essential to notice the factorization of the free wave function, characteristic of the variational method. For the Brueckner correlation function $F_B$ this property does not hold, which embodies the non-locality of the G-matrix. However it can hold approximately, and this can be verified by e.g. the numerical comparison between the two correlation functions. Details on the formal comparison between $F_V$ and $F_B$ can be found in the Appendix.\par
We consider symmetric nuclear matter around saturation and we take the potential Argonne $v_{18}$ \cite{wir} as the two-body nucleon-nucleon interaction. At the Fermi momentum $k_F \,=\, 1.36 $\,fm$^{-1}$, corresponding to density $0.17$ fm$^{-3}$, we compare in Fig. \ref{fig:def1s0} the correlation functions
$F_V(r)$ and $F_B(r)$ at the relative initial momentum $q \,=\, 0.1$ fm$^{-1}$ and at zero total momentum $P$. In this case the correlation functions are calculated for the $^1S_0$ channel. In the variational method a small hard core of radius $R_c \,=\, 0.1 fm$ is introduced for numerical reasons, which is apparent from the figure since the correlation function $F_V$ is zero below the core radius. Both correlation functions feel in any case the repulsive, but finite, core of the interaction and they decrease sharply at short distance. They agree closely above the small core radius $R_c$. At large distances 
both correlation functions reach the expected value of $1$, but just above the repulsive core  
they exceed $1$, due to the attractive part of the NN interaction. In this region they practically coincide. A small discrepancy is observed at intermediate distances, where $F_V$ is slightly larger than $F_B$. To be more quantitative, we calculated the mean absolute deviation for $r > 0.25\,$fm.
We found a value below 2\%, as in all cases we are going to consider in the following. \par 
The comparison for the $^1S_0$ channel, but for $q \,=\, 0.5$ fm$^{-1}$, is reported in Fig. \ref{fig:def1s0bis}. In this case already at moderate distance the two-body wave function $F$ starts to oscillate since it smoothly merges into the free wave function, i.e. the Bessel function (of order $0$ in this case). The same agreement between $F_V$ and $F_B$ is observed. This result indicates that the factorization of 
Eq.(\ref{eq:fv}) is approximately valid also for the correlation function $F_B$ of the BBG expansion.
It is also an indication that the defect function is approximately local. \par
Notice that in the numerical calculations the correlation functions are multiplied by $r^2$ and therefore the contribution of the small distances is vanishing small. This is illustrated in the same figures, where the correlation functions multiplied by $r^2$ are reported. In this case the very close agreement is apparent. A similar trend is obtained for the $^3S_1$ channel, Fig. \ref{fig:def3s1}.    
In the channels with higher partial waves the agreement is even better. To put in evidence the tiny differences, we have reported in an amplified scale the correlation functions in  
Fig. \ref{fig:def1p1} for the $^1P_1$ channel and in Fig. \ref{fig:def3p1} for the $^3P_1$ channel. Notice the change of scale with respect to the previous figures. In these cases the centrifugal barrier suppresses further the two-body wave functions at short and intermediate distance. At larger distance, outside the considered range, the correlation functions merges into the proper Bessel function and then they obviously coincide. \par  
 We also checked the dependence on the total momentum that is present in the two-body wave function.
 It turns out that this dependence is quite weak, see Fig. \ref{fig:def3s1p}, which justifies the assumption, intrinsic in the variational method, of neglecting such a dependence.
 Finally we have introduced the three-body forces (TBF) in the calculations, both in the LOCV and the BBG schemes. It is well known that TBF are necessary if the phenomenological saturation point of nuclear matter has to be reproduced. At the level of two-body correlation approximation, as BHF and LOCV, the TBF are reduced to an effective two-body force by averaging on the position and on spin-isospin of the third particle \cite{andre}. The averaging involves the two-body correlation itself. In principle the original TBF can be derived within the nucleon-meson model of nuclear forces. This procedure turns out to have only a limited success \cite{ZHLi1,ZHLi2} and requires in any case the tuning of the parameters (masses and coupling constants) to get a reasonable saturation point. The latter can be obtained only with the Bonn B potential \cite{BonnB,Adv} as two-body forces \cite{ZHLi1,ZHLi2}.
 We prefer to follow a more pragmatic point of view.  We used the Urbana IX model and treated the TBF according to the method adopted in ref. \cite{lidia}, where the averaging is performed by using a schematic two-body correlation function. We then tune the (two) parameters of  the TBF to get a good saturation point for BHF and we use the same values in LOCV. Around saturation the contribution of TBF is relatively small in absolute value, about 1-3 MeV, in comparison with the total correlation energy that is about -40 MeV at this density. It is slightly repulsive, and as a consequence the two-body wave function is further reduced. This is illustrated in Fig. \ref{fig:def3s1tbf}. Since the effect is quite small, as expected by the relatively weakness of the TBF, Fig. \ref{fig:def3s1tbf}b shows a blow up of the small distance region. It looks that the effect of the TBF is slightly larger for the BHF method.  
 \section{Discussion and conclusion}
We have studied the correlation properties of nuclear matter both in the variational 
LOCV method and in the BHF scheme. In particular we have shown that one can identify the variational (generalized) Jastrow factor $F_V(r)$ with the BHF correlation function $F_B(r) \,=\, 1\,+\,g(r)$, 
where $g(r)$ is the so-called Defect Function. Despite the additional total and relative momentum dependence of $F_B$, not present in $F_V$, and the different method of approximation, it turns out that the two correlation functions are quantitatively quite similar. This is true for each two-body channel, with or without the inclusion of the three-body forces. To see the possible relation of the small differences between the LOCV and BHF correlation functions to other nuclear matter properties, 
we have computed the nuclear matter Equation of State (EOS) in the two theoretical schemes. The results are reported in Fig. \ref{fig:EOS}. The two lower curves, labelled 2BF, correspond to the EOS with two-body forces only, while the two upper curves, labelled 2BF \,+\, 3BF, correspond to the EOS when the (same) three-body forces are also included. One can notice that the two EOS are much more similar when the three-body forces are included. This is in line with the similar finding \cite{shaban} that the EOS' s with different NN interactions become much closer when the (same) three-body force is included. 
\begin{table}[h]
\caption{Nuclear matter correlation energy per particle 
in LOCV and in BHF as a function of the density $\rho$. The first column (K.E.) 
for LOCV gives the modification of the kinetic energy due to the two-body
correlation, 
the second one (P.E.) the potential part and the third one their
sum. For comparison the BHF total correlation energy is reported in the last column
(BHF). The three-body forces are included. }
\begin{tabular}{|c|c|c|}
\hline \hspace*{0.ex} $\rho$(fm$^{-3})$ \hspace*{-0.2ex} & \hspace*{11.6ex}
$LOCV$
\hspace*{10.ex} &$BHF$\hspace*{0.65ex}\\
\hline
\end{tabular}

\begin{tabular}{|c|c|c|c|c|}
\hline \hspace*{8.3ex} & \hspace*{1.5ex} K.E. \hspace*{1.5ex}
& \hspace*{1.5ex} P.E. \hspace*{1.5ex} & \hspace*{1.5ex} TOT
\hspace*{1.5ex} &
\hspace*{2ex}  \hspace*{4ex} \\

\hline

\hspace*{2.ex} 0.10 \hspace*{2.ex}  & 11.24 & -40.75 & -29.51 & -29.24 \\

0.17  & 16.77 & -56.09 & -39.32 & -37.97\\

0.20  & 18.94 & -61.20 & -42.26 & -40.42\\

0.30  & 25.62 & -71.18 & -45.56 & -43.60\\

0.34  & 28.20 & -72.34 & -44.14 & -43.10\\

0.40  & 32.12 & -71.83 & -39.71 & -40.46\\

0.50  & 36.68 & -64.32 & -27.64 & -31.07\\

\hline

\end{tabular}\label{tab_N}
\end{table} 
\par 
In the variational method the average kinetic energy is affected directly by correlations. The total correlation energy includes a kinetic energy part and a potential part, see Eq. (\ref{eq:W}). The breakdown of the two contributions as a function of density is reported in Table I for the case where TBF are included. For comparison the total potential energy of the BHF calculations is also reported. In the BHF scheme the kinetic energy is not explicitly modified \cite{book}, and the whole correlation energy is   
contained in the potential energy coming from the G-matrix contribution. The modification of the kinetic energy is embodied in the momentum dependence of the $G$-matrix and in the self-consistent single particle potential, which also affects the total binding indirectly since it determines the entry energy of the G-matrix. From the results it looks that the connection of the EOS and the details of the correlation function is not so straightforward. 
This is apparent if we calculate the correlations functions at twice the saturation density. They are displayed in Fig. \ref{fig:def1s0_032_tbf}a for the case with only two-body forces. The agreement between the two correlation looks insensitive to the introduction of the TBF, see Fig. \ref{fig:def1s0_032_tbf}b and indeed quantitatively the disagreement, as anticipated before, is below 2\% for $r > 0.25$fm, with or without TBF.
 Despite small variations can be relevant, it looks unlike that
this deviation can be considered responsible of the fact that the disagreement between 
BHF and LOCV is reduced by several MeV at this density once TBF are introduced. It has to be noticed that in BHF there is no  simple way to relate the binding energy to the correlation function, which is not directly involved in the BHF expression for the correlation energy. The change of the binding is clearly due to the direct effect of the change in the nucleon-nucleon force due to the TBF. The only effect of TBF, on the correlation function, see Fig. 
\ref{fig:def1s0_032_tbf}b, seems to be a very small decrease at intermediate distance of $F_V$ with respect to $F_B$. This could suggest that the good agreement of the EOS is the result of a redistribution of the attractive and the repulsive contributions to binding. To make easier the qualitative estimate of the relevance of the TBF, we have reported in Fig. \ref{fig:016_032} the comparison of the correlation functions with and without TBF at the densities 0.16 fm$^{-3}$ and 0.32 fm$^{-3}$, both for LOCV and BHF. At increasing density the effect of TBF increases, but the effect looks larger for BHF. Also in this case no systematic trend is observed in relation to the corresponding EOS. It has been found in ref. \cite{Bozek} that also the spectral function has a mild dependence on the presence of TBF.        
Beside the EOS, other quantities, like transport coefficients or neutrino and electron scattering cross sections, are probably more directly related to the correlation and spectral function
\cite{Umberto}. The analysis of this point is left to a future work, but in any case no major discrepancy can be expected between BHF and LOCV schemes.      
 
\begin{acknowledgments}
One of the authors (HRM) would like to thank
University of Tehran for partially supporting him under the grants
provided by research council. This work was supported by INFN Sezione di Catania within the national project CT51.
\end{acknowledgments}
\appendix*

\section{}

In this Appendix we give some details on the formal comparison between the effective correlation factor $F_B$ that is present in the ground state wave function of the Brueckner approximation, within the BBG hole-line expansion, and the corresponding correlation factor $F_V$ in the variational LOCV approximation. The unperturbed ground state $\Phi$ is the anti-symmetrized product of N single particle momentum states 
\begin{equation}
\vert \Phi \ket \,=\, \Pi_{k_i} a^\dagger_{k_i} \vert O \ket , 
\end{equation}  
\noindent
where $\vert O \ket$ is the vacuum state and the $k_i$ include spin-isospin variables. 
Then Eq. (\ref{eq:prod}) can be rewritten 
\begin{equation}
\begin{array}{ll}
\vert \Psi \ket \, &=\,\, \Pi_{\{k_1 k_2\} } \hat{F}_{k_1 ,k_2 } \vert O \ket  \\ 
     \    &\,                    \\ 
 \hat{F}_{k_1 k_2} \, &=\,\, \sum_{k_1' k_2' } \left[ \delta_{k_1' k_1} \delta_{k_2' k_2} \,+\,
 \bra  k_1' k_2' \vert Q S_2 \vert k_1 k_2 \ket \right] {a}\creas(k_1'){a} \creas(k_2') ,         
\end{array}
\end{equation}
\noindent
where the summations in $\hat{F}$ are over all momenta and we have introduced the Pauli operator $Q$ that restricts the momenta $k_1' k_2'$   
outside the Fermi sphere, while the momenta $k_1 k_2$ are inside the Fermi sphere. 
\par
It is convenient to introduce the wave function of the correlated ground state $\vert \Psi \ket$
by taking the scalar product with the anti-symmetrized N-particle coordinate states 
\begin{equation}
\vert r_1 r_2 .... r_N \ket \,=\, \Pi_i \psi^\dagger (r_i) \vert O \ket \,=\, \vert \{r_i\} \ket 
\end{equation}
\noindent where $\psi^\dagger (r_i)$ is the creation operator of a particle at the position $r_i$
(including spin-isospin variables).
One gets
\begin{equation}
 \Psi(\{r_i\}) \,=\,  \bra \{r_i\} \vert \Psi \ket \,=\, A_{\{r_i\}} \Pi_{\{k_1 k_2\}} 
 f_{k_1 k_2 }(r_i,r_j)    
\end{equation}
\noindent where the operator $A$ anti-symmetrizes the N coordinates $r_i$ and
\begin{equation}
 f_{k_1 k_2 }(r_i,r_j) \,=\, \sum_{k_1' k_2' } \left[ \delta_{k_1' k_1} \delta_{k_2' k_2} \,+\,
 \bra  k_1' k_2' \vert Q S_2 \vert k_1 k_2 \ket \right]\bra r_i \vert k_1'\ket \bra r_j \vert k_2' \ket,
\end{equation}
\noindent which is the Fourier transform of the defect function. The variables $r_i$ and $r_j$ are two generic coordinates among the N anti-symmetrized ones.  Introducing the coordinate representation for the defect function, one gets
\begin{equation}
\begin{array}{ll} 
f_{k_1 k_2 }(r_i,r_j) &\,=\, \bra r_i \vert k_1\ket \bra r_j \vert k_2 \ket \,+\, \bra r_i r_j \vert 
Q S_2 \vert k_1 k_2 \ket \\
  \  &\,   \\
  \  &\,=\, \bra r_i \vert k_1\ket \bra r_j \vert k_2 \ket \,+\, \int d^3r_i'd^3r_j' 
 \,  \bra r_i r_j \vert Q S_2 \vert r_i' r_j' \ket \, \bra r_i' \vert k_1 \ket \bra r_j' \vert k_2 \ket. \\  
\end{array} 
\end{equation}
\noindent We consider the relative coordinate $r_{ij} = (r_i - r_j)$ and center of mass coordinate 
$R_{ij} = (r_i + r_j)/2$ and notice that the defect function $QS_2$ is diagonal in the total momentum $P$. If furthermore we assume that the defect function is local, one gets
\begin{equation}
\begin{array}{ll} 
f_{k_1 k_2 }(r_i,r_j) &\,=\, \bra r_{ij} \vert q \ket \bra R_{ij} \vert P \ket \,+\,
\int d^3 r_{ij}' \bra r_{ij} \vert QS_2(P) \vert r_{ij}' \ket \,\bra r_{ij}' \vert q \ket \bra R_{ij} \vert P \ket \\
 \  &\,  \\
 \  &\,=\, \left[ 1 \,+\, g(r_{ij}) \right] \bra r_{ij} \vert q \ket \bra  R_{ij} \vert P \ket, \\ 
\end{array} 
\end{equation}
\noindent where $q$ is the relative momentum and 
\begin{equation}
\bra r_{ij} \vert QS_2(P) \vert r_{ij}' \ket \,=\, g(r_{ij}) \delta(r_{ij} - r_{ij}').
\end{equation}
\noindent Here the dependence on the total momentum of the defect function has been neglected.
This result shows that, under the stated assumptions, the correlated wave function can be written as
\begin{equation}
\Psi(\{r_i\}) \,=\,A\Pi_{k_1 k_2} \left[ 1 \,+\, g(r_{ij}) \right] \bra r_i \vert k_1 \ket \bra r_j \vert k_2 \ket
\end{equation}
\noindent which has the form of the variational wave function, if we identify the factor $1 + g$ with the correlation function $f(r)$ of the variational method. The defect function in the mixed representation $F_B$ has then to be compared
with $F_V(r) \bra r \vert q \ket$, as discussed in the text. Both $F_B$ and $F_V$ can be expanded in partial waves and compared channel by channel.

\vfill\eject
\begin{figure}[t]
\vskip -10 cm
\begin{center}
\includegraphics[bb= 140 0 300 790,angle=0,scale=1.]{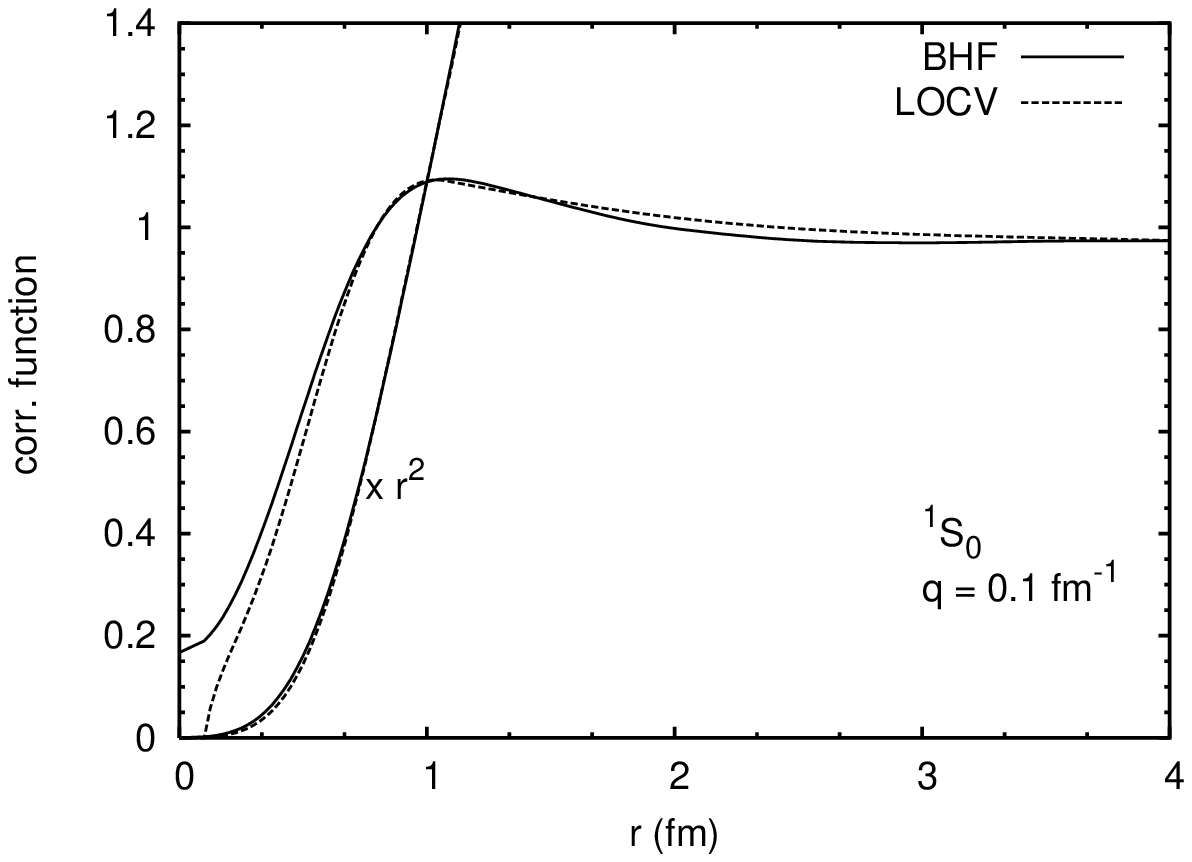}
\vskip -1 cm
\caption{Correlation function in the $^1S_0$ channel for the LOCV and BHF approaches. The same correlation functions multiplied by $r^2$ are also shown. The momentum $q = 0.1$fm$^{-1}$ is the
relative momentum of the two correlated particles.} 
\label{fig:def1s0}
\end{center}
\end{figure}
\begin{figure}[t]
\vskip -12 cm
\begin{center}
\includegraphics[bb= 140 0 300 790,angle=0,scale=1.]{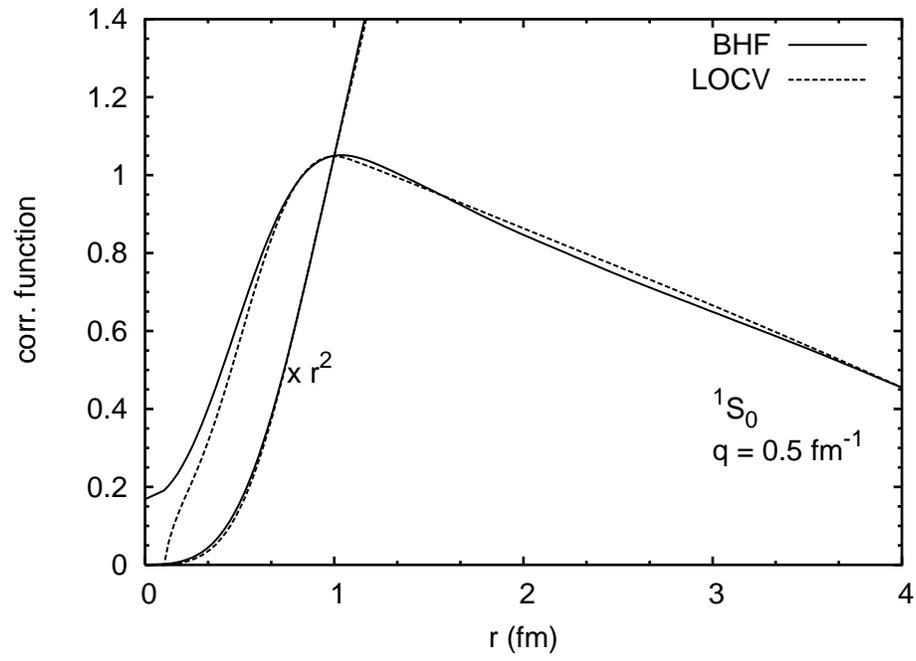}
\vskip -0.5 cm
\caption{The same as in Fig. \ref{fig:def1s0}, but for $q = 0.5$fm$^{-1}$.} 
\label{fig:def1s0bis}\end{center}
\end{figure} 
\vskip 2 cm

\vfill\eject

\begin{figure}[t]
\vskip -12 cm
\begin{center}
\includegraphics[bb= 140 0 300 790,angle=0,scale=1.]{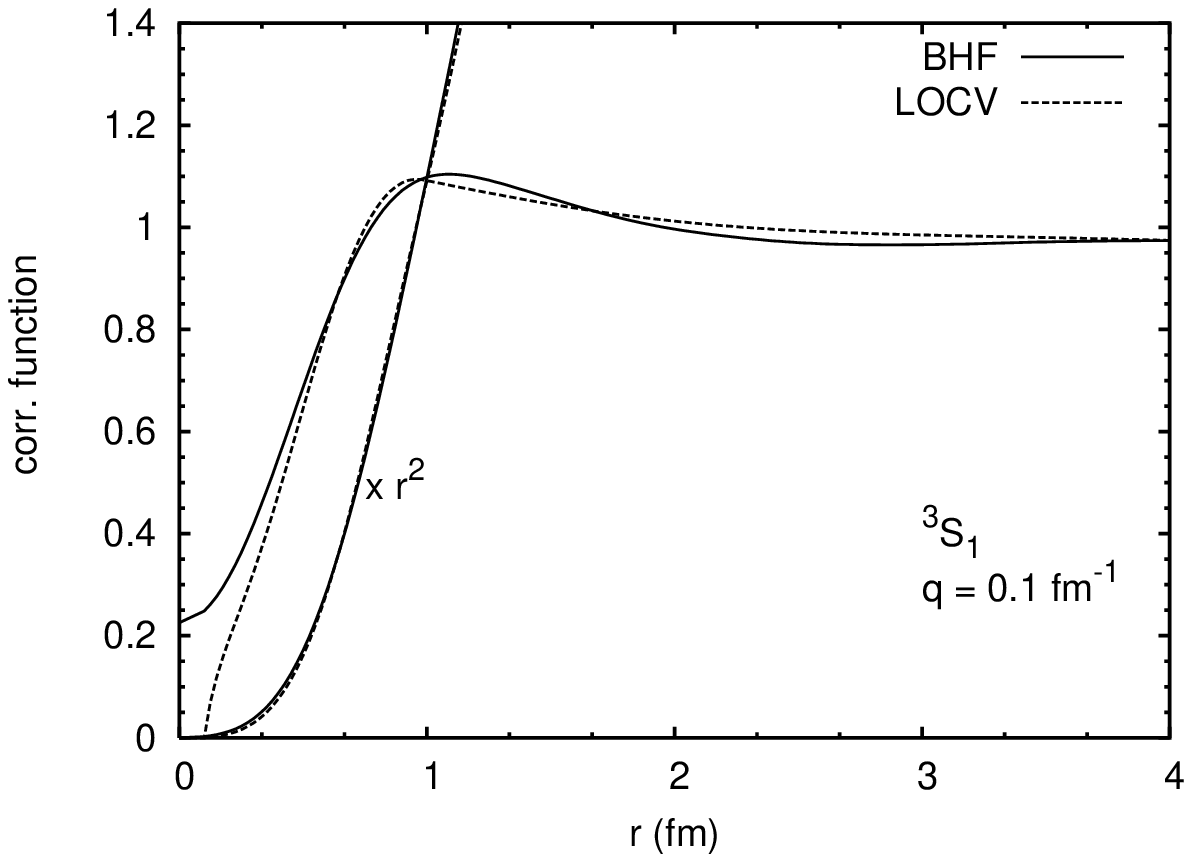}
\vskip -0.5 cm
\caption{The same as in Fig. \ref{fig:def1s0}, but for the $^3S_1$ channel.} 
\label{fig:def3s1}\end{center}
\end{figure} 
\vskip 2 cm

\vfill\eject

\begin{figure}[t]
\vskip -12 cm
\begin{center}
\includegraphics[bb= 140 0 300 790,angle=0,scale=1.]{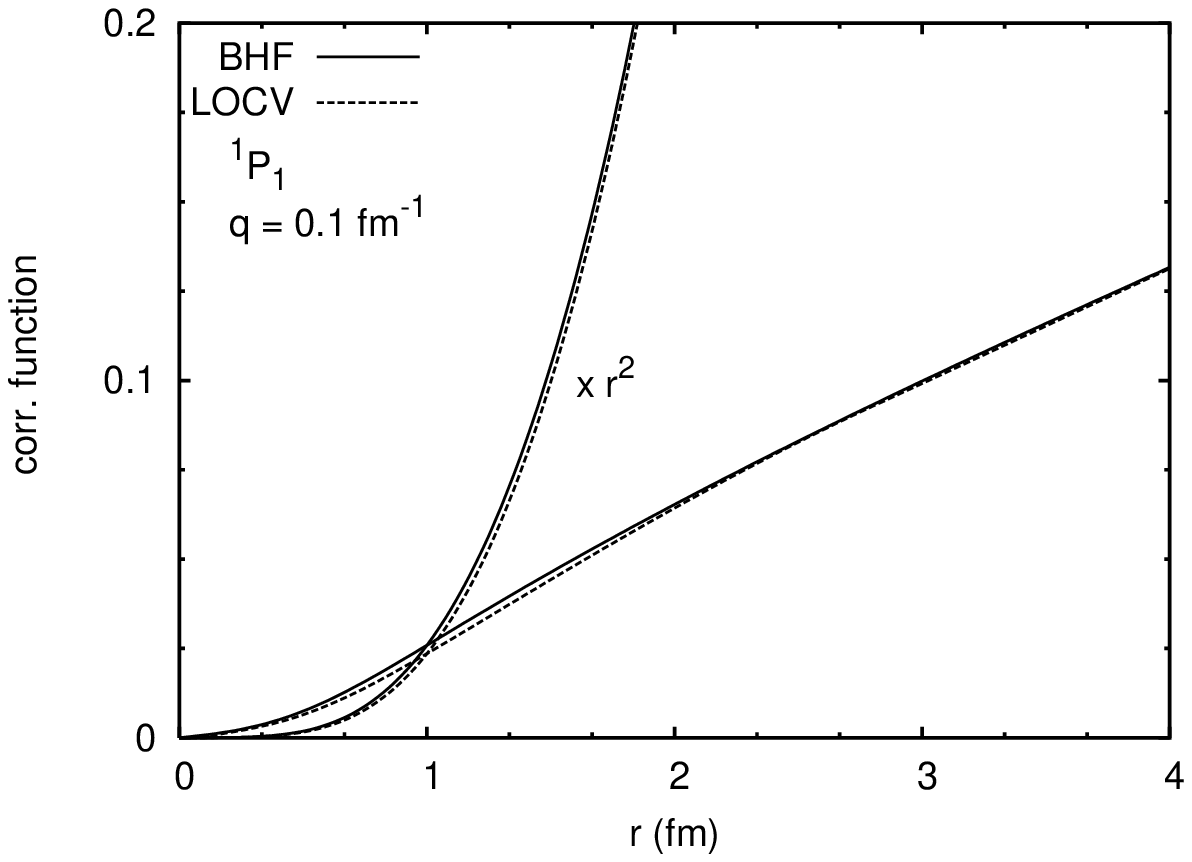}
\vskip -0.5 cm
\caption{The same as in Fig. \ref{fig:def1s0}, but for the $^1P_1$ channel.} 
\label{fig:def1p1}\end{center}
\end{figure} 
\vskip 2 cm

\vfill\eject

\begin{figure}[t]
\vskip -12 cm
\begin{center}
\includegraphics[bb= 140 0 300 790,angle=0,scale=1.]{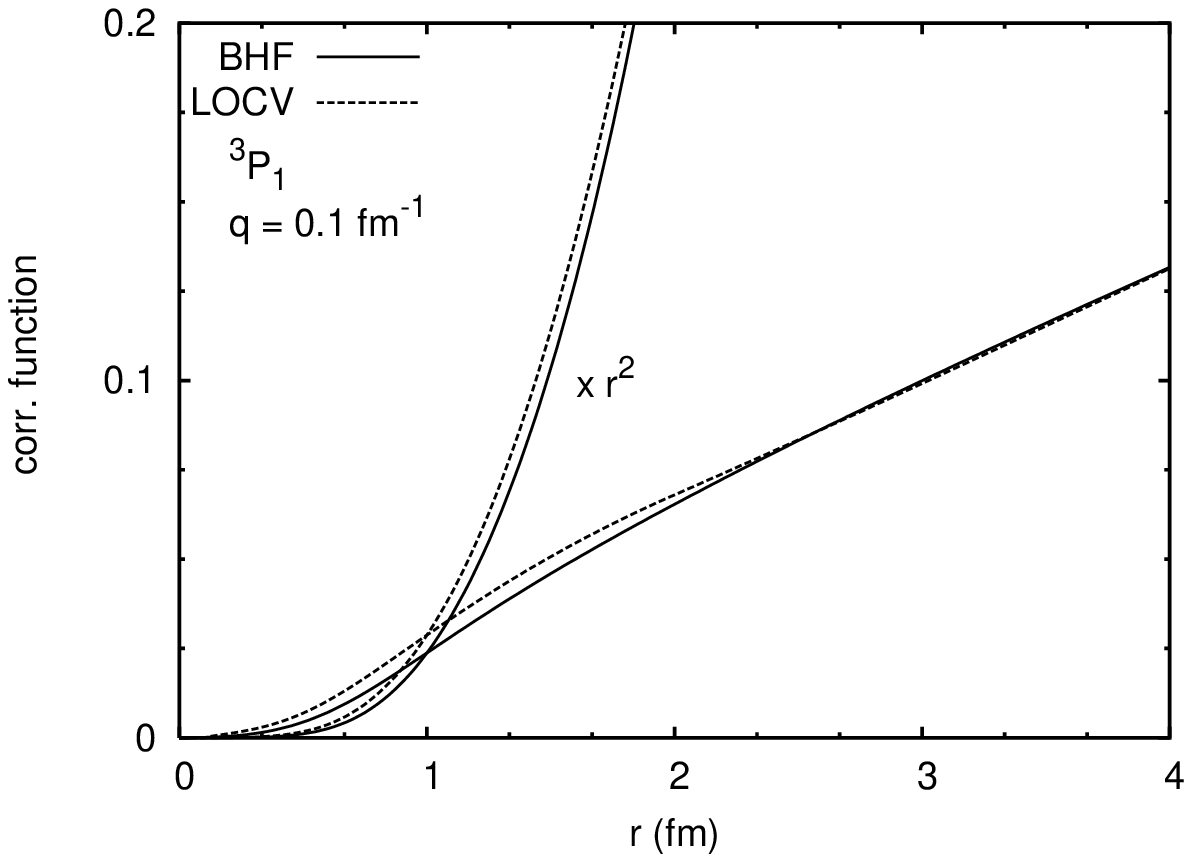}
\vskip -0.5 cm
\caption{The same as in Fig. \ref{fig:def1s0}, but for the $^3P_1$ channel.}
\label{fig:def3p1}\end{center}
\end{figure} 
\vskip 2 cm

\vfill\eject

\begin{figure}[t]
\vskip -12 cm
\begin{center}
\includegraphics[bb= 140 0 300 790,angle=0,scale=1.]{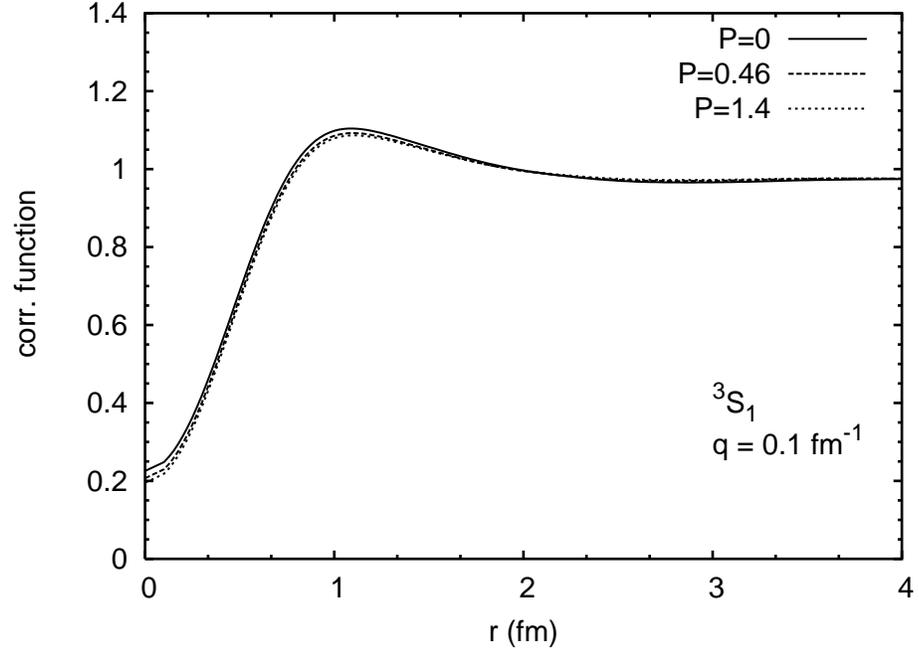}
\vskip -0.5 cm
\caption{The correlation functions as in Fig. \ref{fig:def3s1} at different total momentum $P$ of the two correlated particles.}
\label{fig:def3s1p}\end{center}
\end{figure} 
\vskip 2 cm

\vfill\eject
\begin{figure}[t]
\vskip -12 cm
\begin{center}
\includegraphics[bb= 140 0 300 790,angle=0,scale=0.8]{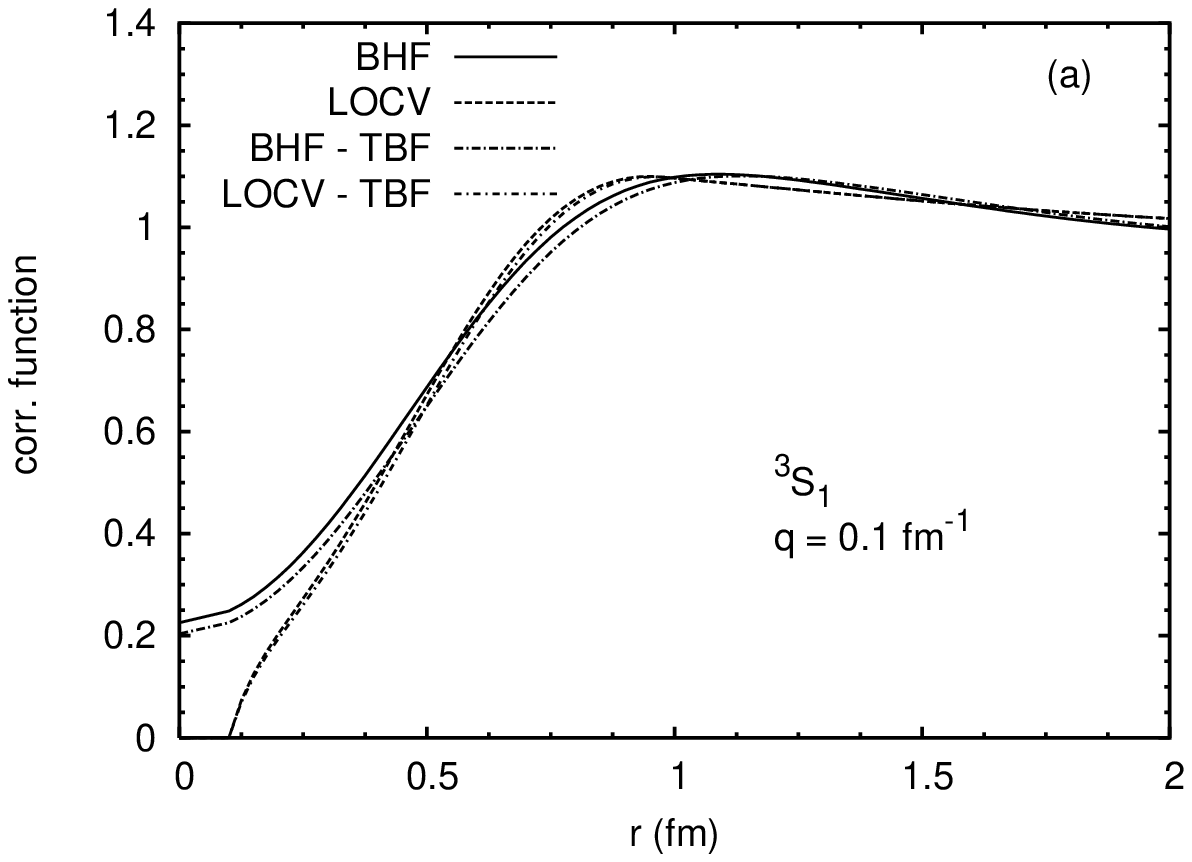}
\vskip -0.5 cm
\end{center}
\end{figure} 
\begin{figure}[t]
\vskip -28 cm
\begin{center}
\includegraphics[bb= 140 0 300 790,angle=0,scale=0.8]{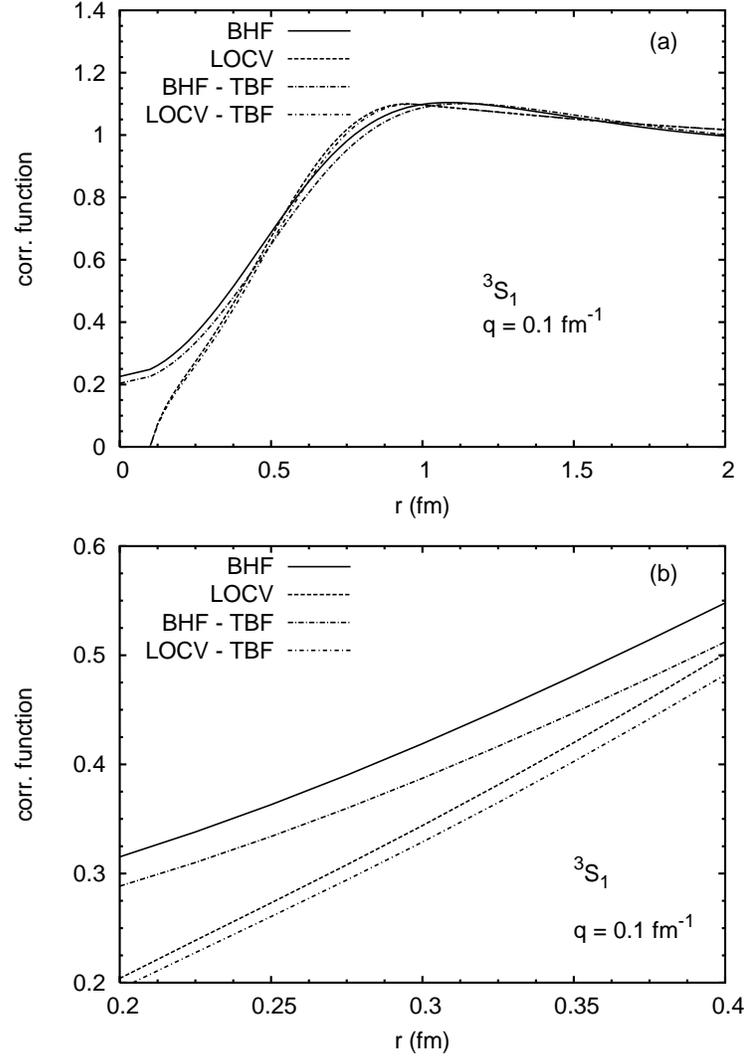}
\vskip -1.5 cm
\caption{In panel (a) the correlation function in the $^3S_1$ channel is reported, with
and without three-body forces.  Panel (b) is the blow up of the plot in panel (a)
within a region at small distances.} 
\label{fig:def3s1tbf}\end{center}
\end{figure} 

\vfill\eject

\begin{figure} [t]
\vskip -12 cm
\begin{center}
\vskip 0.6 cm
\includegraphics[bb= 140 0 300 790,angle=0,scale=1.0]{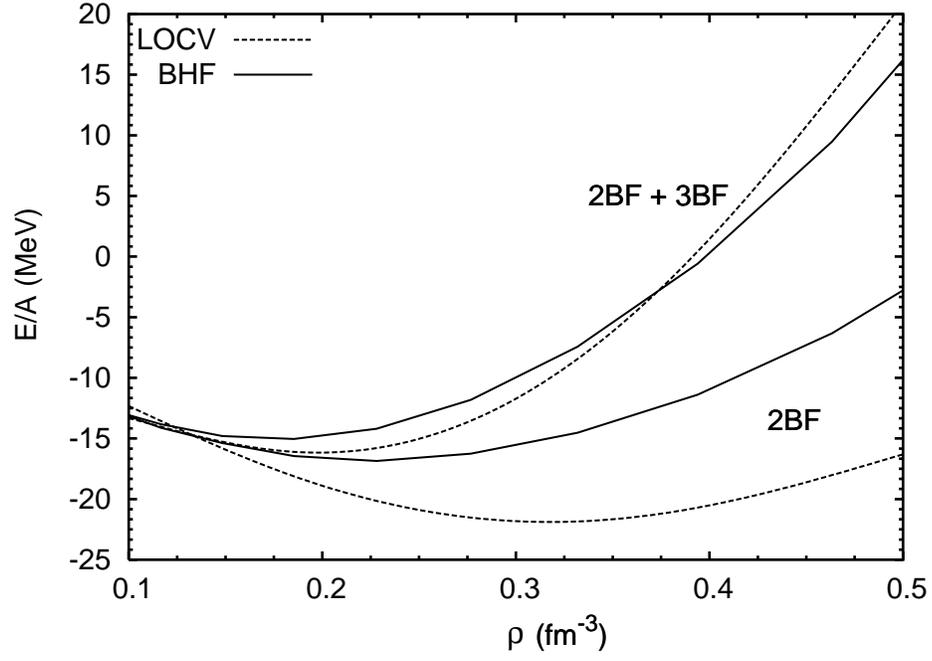}
\vspace{0.3 cm}
   \caption{ Binding energy per particle as a function of the density $\rho$ in symmetric nuclear matter for the LOCV and BHF approaches. The two lower curves, labelled 2BF correspond to calculations with two-body force only. The two upper curves, labelled 2BF + 3BF correspond to calculations with the inclusion of the three-body forces. }   
\label{fig:EOS}\end{center}
 \end{figure}

\vfill\eject
\begin{figure}[t]
\vskip -10 cm
\begin{center}
\includegraphics[bb= 140 0 300 790,angle=0,scale=1.]{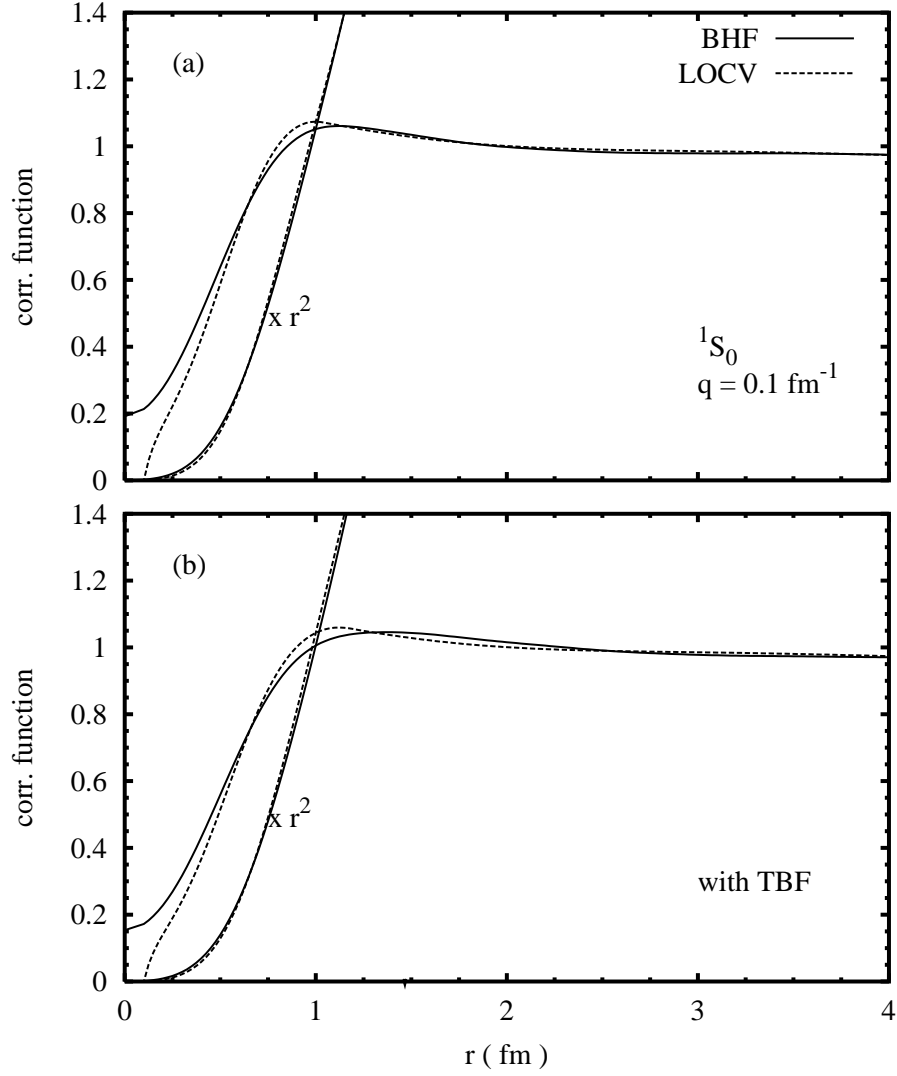}
\vskip -1 cm
\caption{Correlation function in the $^1S_0$ channel for the LOCV and BHF approaches at the density $\rho \,=\, 0.32$ fm$^{-3}$. The same correlation functions multiplied by $r^2$ are also shown. The momentum $q = 0.1$fm$^{-1}$ is the
 relative momentum of the two correlated particles. Panel (a) : only two-body forces. Panel (b) : also three-body forces are included.}
\label{fig:def1s0_032_tbf}\end{center}
\end{figure}

\vfill\eject
\begin{figure}[t]
\begin{center}
\includegraphics[bb= 340 0 440 790,angle=0,scale=0.7]{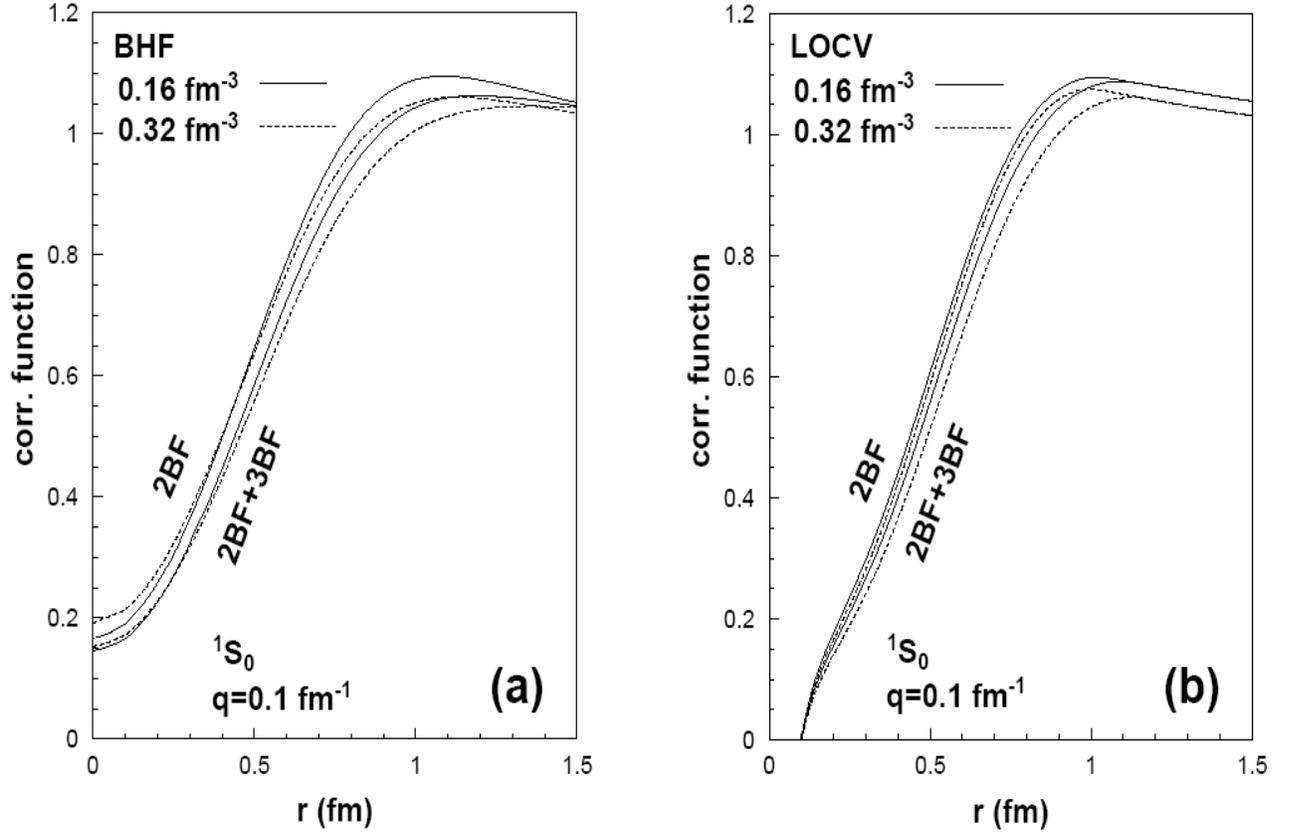}
\vskip -1 cm
\caption{ Comparison of the effect of the TBF on the correlation function at two different densities,
0.16 fm$^{-3}$ (full lines) and 0.32 fm$^{-3}$ (dashed lines). At each density the lower curves include the TBF. Panel (a) refers to BHF, panel (b) to LOCV. The meaning of the other labels is as in previous figures.}
\label{fig:016_032}\end{center}
\end{figure}

\end{document}